\documentclass[preprint,12pt]{aastex}

\newcommand{\ha}{H\ensuremath{\alpha}}
\newcommand{\hb}{H\ensuremath{\beta}}

\newcommand{\nii}{[N\,II]}
\newcommand{\sii}{[S\,II]}
\newcommand{\oiii}{[O\,III]}

\begin{document}
\title{Difference in Narrow Emission Line Spectra of Seyfert 1 and 2 galaxies}

\author{
Kai Zhang, Tinggui Wang, Xiaobo Dong, Honglin Lu}
\affil{Center for Astrophysics, University of Science and
Technology of China (USTC), Hefei, Anhui, 230026, China; ~
zkdtc@mail.ustc.edu.cn, twang@ustc.edu.cn}

\affil{Joint Institute of Galaxies and Cosmology, USTC and Shanghai
Observatory}

\shorttitle{Narrow Line Spectra of Seyfert galaxies}
\shortauthors{Zhang et al.}

\begin{abstract}
In the unification scheme of Seyfert galaxies, a dusty torus blocks
the continuum source and broad line region in Seyfert 2 galaxies.
However it is not clear whether or not and to what extent the torus
affects the narrow line spectra. In this paper, we show that Seyfert
1 and Seyfert 2 galaxies have different distributions on the
[O\,III]/H$\,\beta $ vs [N\,II]/H$\,\alpha$ diagram (BPT diagram)
for narrow lines. Seyfert 2 galaxies display a clear left boundary
on the BPT diagram and only 7.3\% of them lie on the left. By
contrast,  Seyfert 1 galaxies do not show such a cutoff and 33.0\%
of them stand on the left side of the boundary. Among Seyfert 1
galaxies, the distribution varies with the extinction to broad
lines. As the extinction increases, the distribution on BPT diagram
moves to larger [N\,II]/H$\,\alpha$ value. We interpret this as an
evidence for the obscuration of inner dense narrow line region by
the dusty torus. We also demonstrate that the [O\,III] and broad
line luminosity correlation depends on the extinction of broad lines
in the way that high extinction objects have lower uncorrected
[O\,III] luminosities, suggesting that [O\,III] is partially
obscured in these objects. Therefore, using [O\,III] as an indicator
for the nuclear luminosity will systematically under-estimate the
nuclear luminosity of Seyfert 2 galaxies.

\end{abstract}




\section{Introduction}

The unification scheme of Seyfert galaxies has been widely accepted
to explain the dichotomy of Seyfert 1 (hereafter Sy1) and Seyfert 2
galaxies (hereafter Sy2) (e.g. Antonucci 1993). The basic idea is
that the differences between Sy1s and Sy2s are caused by an
orientation effect. The broad line region (BLR) and continuum source
are blocked from our view in Sy2s by thick obscuring material,
presumably a dusty torus (Antonucci \& Miller 1985) at large
inclination angles, while they are observed directly in Sy1s. In
both types of Seyfert galaxies, narrow emission lines, produced on a
large scale, are observed. This scheme has been supported by many
observations (Antonucci \& Miller, 1985; Miller \& Goodrich, 1990;
Kl\"{o}ckner et al, 2003; Greenhill et al, 2003; Jaffe et al, 2004),
and even the temperature distribution  of the dusty torus has been
measured (e.g., Tristram et al, 2007).

  While it is general agreed that dusty tori obscure BLR and
continuum source, and even the central narrow line region (NLR) of
Sy2s (e.g., Rhee \& Larkin 2005; Haas et al. 2005), it is still
unclear to what extent they obscure the NLR, and how the obscuration
depends on the orientation (i.e., the difference in NLR obscuration
among different types of Seyfert galaxies). This is an important
issue because the narrow line luminosity is often used as a
surrogate for the bolometric luminosity of Seyfert 2 nuclei. In
addition, it may yield important clues to the structure of NLR.
Jackson \& Browne (1990) compared [O\,III] emission from quasars
(i.e., radio-loud QSOs) with radio galaxies --- both of them are
believed to be the analogs of the two types of Seyfert galaxies
(Barthel 1989), and found that the [O\,III] emission of quasars is
much stronger than that of radio galaxies. It was interpreted as
part of the [O\,III] emission is obscured by the torus in radio
galaxies. A similar conclusion was reached by Haas et al. (2005) by
comparing [O\,III]$\lambda$5007 with [OIV]$\lambda$25.9$\mu$m. On
the other hand, Barthel \& Fosbury (1994) showed that, when
comparing [OII] emission of quasars and radio galaxies, both classes
of objects have very similar distributions, corroborating the
obscuration scenario. Unfortunately, [OII] emission may be enhanced
by shocks induced by radio plasma (Best et al. 1999) and thus may
not be a good tracer of AGN luminosity. Moreover, a fraction of FRII
radio galaxies may be intrinsically less luminous than quasars
(Ogle, Whysong \& Antonucci, 2006; also Meisenheimer et al. 2001).

It is well known that NLR is stratified, at least for AGNs of some
sort (e.g.,Veilleux et al. 1991, Robinson et al. 1994). Emission
lines of higher critical density or ionization potential are
produced averagely closer to the active nucleus than those of lower
critical density or ionization. This fact can be used to check
whether the inner NLR region is blocked by torus in Sy2s; and if
yes, to what extended NLR is obscured. Murayama \& Taniguchi (1998)
found that Sy1s show excess [Fe VII]$\lambda$6087 with respect to
Sy2s, and proposed that [Fe VII] arises mainly from the inner wall
of dusty tori. Schmitt (1998) compared [OII]$\lambda$3727\AA,
[NeIII]$\lambda$3869\AA, [O\,III]$\lambda$5007\AA,
[NeV]$\lambda$3426\AA, as well as 60$\mu$m continuum flux from a
sample of 52 Sy1s to those of 68 Sy2s; they found that Sy1s have
higher excitation than Sy2s. However, the differences were
interpreted as that Sy1s have smaller number of ionization bounded
clouds than Sy2s by him.

In this paper, we study the location of Sy1s and Sy2s on the BPT
diagrams, short for Baldwin-Phillips-Terlevich diagrams (Baldwin et
al 1981), which are often used for classification of narrow emission
line galaxies. With the large spectroscopic sample of both Sy1s and
Sy2s available from the Sloan Digital Sky Survey (York et al, 2000),
and spectral decomposition, we are able to detect the different
distributions of Sy1s and Sy2s on BPT diagrams.

\section{Sample and Data Analysis}
\subsection{Seyfert 1 and 2 galaxy samples}

The Seyfert 2 galaxies sample is selected from the galaxy catalog of
SDSS Data Release 4 (DR4). For our purpose, a redshift cut of
z$<$0.3 is used to ensure that  the emission lines of interest,
H$\,\alpha$, H$\,\beta$ , [O\,III] and [N\,II], [S\,II], fall in the
spectra. We subtracted the stellar continuum to leave a clean
emission line spectrum following the recipe described in  Lu et al.
(2006). In brief, we fit galaxy spectra with the templates derived
by applying Ensemble Learning for Independent Component Analysis
(EL-ICA) to the simple stellar population library (Bruzual \&
Charlot 2003). The templates were then broadened and shifted to
match the stellar velocity dispersion of the galaxy. In this way,
the stellar absorption lines are reasonably well subtracted to
ensure reliable measurement of weak emission lines. Next, we fitted
the emission line spectrum using gaussians (see next section) to
derive emission line parameters. In practice, stellar subtraction
and the emission line fitting are iterated because line parameters
are used to create correct mask regions for emission lines during
the stellar modeling. Only sources with H$\alpha$, H$\beta$ ,
[O\,III], [N\,II] and [S\,II] lines detected with S/N $> 5$ are
considered for further study. Broad emission line objects are
removed from the Seyfert 2 galaxies sample (described below). We
adopt the criteria of Kewley et al. (2006) to classify the emission
line galaxies as star forming galaxies, Seyfert galaxies, LINER,
AGN/HII composites. For this study, we only stick to the classical
Seyfert galaxies because only a very small fraction of Sy1s show a
composite/LINER spectrum while there are large portion of such Sy2s.
The final sample consists of 5544 sources.

Seyfert 1 galaxies\footnote{A small number of objects are in the
luminosity range for quasars.} were drawn from quasar and galaxy
catalogues of SDSS DR4 at redshift $z<0.3$. For nucleus dominated
sources where Fe\,II multiplets and other broad emission lines are
highly blended, we fit simultaneously the nuclear continuum, the
Fe\,II multiplets and emission lines (see Dong et al. 2008 for
details). The nuclear continuum is approximated by a broken
power-law, with free indices for \ha\ and \hb\ region respectively.
Fe II emission, both broad and narrow, is modeled using the spectral
data of the Fe II multiplets for I Zw 1 provided by Veron-Cetty et
al. (2004). Emission lines are modeled as multiple gaussians. For
those spectra with significant contribution of starlight as measured
by the equivalent widths (EWs) of the Ca\,II\,K $\lambda3934$
absorption line or high order Balmer lines or Na\,I
$\lambda\lambda5890,5896$, a starlight model is also included using
the 6 IC templates as described in the above. The decomposition of
host-galaxy starlight, nuclear continuum and FeII emission were
carried out following the procedure as described in detail in Zhou
et al. (2006). We fit and subtracted the above model from the SDSS
spectra, and then fit emission lines using multi-gaussian model (see
next section).

  Based on the fitted emission-line parameters, we construct
the sample of Sy1s according to the criteria as follows. (1) Adding
a broad gaussian component of H$\alpha$ to the model can improve the
fit significantly with a chance probability less than 0.05 according
to F-test; (2) The broad component is detected with S/N $\gtrsim 5$;
(3) The peak of the broad component is at least twice the
root-mean-square (rms) deviation in the regions surrounding \ha\
(Dong et al., in prep). To guarantee the reliability of the
measurements of the narrow lines of interest, we require that they
have S/N $> 5$, the same as in the cases of Sy2s. Here the flux
errors of narrow lines, especially \ha\ and \hb\ narrow lines, have
taken account for the uncertainty introduced by the emission-line
modeling (Dong et al., in preparation). The final sample has 2768
Sy1s.

\subsection{Emission Line Fitting}

For Sy2s, after removing the starlight, we fit the emission-line
spectra using the code described in detail in Dong et al. (2005).
Briefly, every line is fitted with one or more Gaussians as is
statistically justified (mostly with 1--2 Gaussians); the line
parameters are achieved by minimizing $\chi^2$. The
\oiii$\lambda\lambda4959,5007$ doublet lines are assumed to have the
same profiles and redshifts; \nii$\lambda\lambda6548,6583$ and
\sii$\lambda\lambda6716,6731$ doublet are constrained in the same
way. Furthermore, the flux ratios of \oiii\ doublet and \nii\
doublet are fixed to the theoretical values. Usually, \ha\ line and
\nii\ doublet are highly blended and thus hard to isolate; in such
cases, we fit them assuming they have the same profile as \sii\
doublet, which is empirically justified (e.g., Filippenko \& Sargent
1988; Ho et al. 1997; Zhou et al. 2006).

For Sy1s, once the nuclear continuum and the Fe\,II emission are
subtracted, we perform a refined fit of the emission-line spectra.
The narrow lines are modeled as for Sy2s. The broad lines are
modeled with multi-gaussians, as many as is statistically justified.
One concern is the reliability of the decomposition of narrow and
broad lines. We found that when the broad line is significantly
broader than the narrow line and the narrow line is not weak, the
flux of narrow line is trustworthy (Dong et al., in preparation).
This is usually the case for narrow lines (S/N$>5$) of Sy1s in this
sample.

\section{Results}

\subsection{A Comparison between Seyfert 1 and 2 galaxies}

Dong et al.(2008) found that the distribution of broad-line Balmer
decrement for blue AGNs is a gaussian in the logarithmic space, with
a peak at 0.486 (H$\,\alpha$/H$\,\beta$=3.1) and an intrinsic
standard deviation only $\sim$0.03. The reddening interpretation of
large Balmer decrements is also supported by larger infrared to
broad line ratio (Dong et al. 2005) and excess X-ray absorption in
those objects (Wang et al. 2008). We estimate reddening to broad
lines for objects with $H\alpha/H\beta$ significantly larger than
the above value assuming a SMC extinction curve, by
$E_{B-V}^{b}=1.99\times [\log (H\alpha/H\beta)-0.486]$. Extinction
to narrow lines is also estimated using the Balmer decrement and
assuming an intrinsic narrow-line Balmer decrement of 3.1. Using the
$E_{B-V}$ and extinction curve, the intrinsic luminosities of
H$\,\alpha$, H$\,\beta$ broad lines, as well as H$\,\alpha$,
H$\,\beta$, [O\,III] [N\,II], [S\,II] narrow lines can be
calculated.

We plot the 5544 Sy2 and 2768 Sy1 sources on the BPT diagram in
Figure 1. The intermediate broad line $E_{B-V}^{b}$ group are not
plotted for clarity. One can easily see that Sy1s and Sy2s
apparently occupy different regions on BPT diagram. We defined a
line (S12 line for short) such that Sy2s rarely appear on the left
side:
\begin{equation}
\log([O\,III]/H\beta)= 3.53\times \log([N\,II]/H\alpha)+1.65
\end{equation}
Only 409 (7.3\%) Sy2s lie on the left side of S12 line while 913
(33.0\%) Sy1s on the left side. The overall trend is very clear that
Sy1s lie to the left of Sy2s on the diagram. On average,
[N\,II]/H$\alpha$ ratio is 0.11 dex higher in Sy2s than in Sy1s.

\subsection{Seyfert 1 galaxies of different extinction}
If we regard Sy2s as Seyfert galaxies with extremely high
$E_{B-V}^{b}$ in broad lines, the trend of different types of
Seyfert galaxies should also be traced for Sy1s of different broad
line extinction $E_{B-V}^b$. In the right panel of Fig 1, we split
the source according to their $E_{B-V}^{b}$ of broad emission lines.
Only Sy1s in $E_{B-V}^{b}\in[0, 0.2]$ and $E_{B-V}^{b}\in[0.6, 1]$
are plotted for clarity. It is evident that as the $E_{B-V}^{b}$
increases, the distribution moves to right. This is in accordance
with our hypothesis. One can see that a small fraction (23.8\%) of
objects in the $E_{B-V}^{b}\in[0.6, 1]$ group lie on the left side
of S12 line while the fraction for the low extinction group is
(43.3\%).

\subsection{On the narrow and broad line correlation}

From the broad line sample, an empirical relationship between
uncorrected [O\,III] luminosity and broad H$\,\alpha$ line
luminosity can be obtained. The broad H$\,\alpha$ is corrected from
the extinction to the broad line region as in last section.
\begin{equation}
\log(L_{[O\,III]}^{uncorrected} )=(0.977\pm0.007)\times
log(L_{H\,\alpha^b}^{intri} )+(0.238\pm0.294)
\end{equation}
In Fig 2, we split the sample by their $E_{B-V}^{b}$ and plot them
on the [O\,III] vs H$\,\alpha$ diagram. The purple crosses are with
low $E_{B-V}^{b}\in[0,0.2]$ while green triangle symbols with high
$E_{B-V}^{b}\in[0.6,1]$. The average ratios (in logarithmic value)
of luminosities of [O\,III] and H$\,\alpha$ for these two groups are
$-0.678\pm0.278$ and $-1.012\pm0.158$. Thus the [O\,III]
luminosities of the low $E_{B-V}^{b}$ group are on average two times
(0.334 dex) larger than that of high $E_{B-V}^{b}$ group at a given
broad line luminosity. The mean values of the two groups are
significantly different at a chance probability less than $10^{-10}$
according to the Student's t-test.


One concern is that this may be introduced by the biases that the
measurement error in the $E_{B-V}^{b}$ will lead to a shift in the
relation for different $E_{B-V}^{b}$. However, Monte-Carlo
simulation shows that it causes a deviation of only 0.06 dex for a
typical H$\,\alpha$ luminosity, much smaller than the difference of
0.334 dex. Thus this must be real.

\section{Discussion and conclusion}

  We find that Seyfert 1 galaxies have significantly smaller
[N\,II]/H$\alpha$ ratios than Seyfert 2 galaxies. A similar trend
has also been observed in Sy1s with different BLR reddening:
unreddened Seyfert 1 galaxies have smaller ratios than reddened
ones. We also show that reddened Seyfert 1 galaxies have average
lower uncorrected [O\,III] luminosity at a given nuclear luminosity
represented by broad H$\alpha$ luminosity.

Before discussing the implication of these findings, we would like
to rule out the possibility that they are introduced by systematic
bias in our spectral modeling and sample selection. First, as a
sanity check, we fit all the above-mentioned narrow lines assuming
that they have the same profile; it does not change our results.
Thus our findings are not caused by the emission-line modeling
(different models for different lines). Second, stellar absorption
lines are properly subtracted in both Sy1s and Sy2s. We have
visually inspected the high-order Balmer absorption lines to ensure
they are well matched. Thus we are confident that the fluxes of
narrow lines are not affected seriously by stellar absorption
features. Third, we check Sy1s with either small difference in line
width between broad lines and narrow lines, e.g., narrow line Sy1s,
or with a relative low signal to noise ratio, where the
decomposition of narrow and broad components is most difficult. We
find that their location on BPT diagram is indistinguishable from
other Sy1s. Therefore, the line-profile decomposition should not
affect our results. Last, but not the least, we check the influence
of our sample selection by raising the criterion for the S/N of
narrow line to be 7. We find that the fraction of objects lying on
the left side of S12 line is 9\% for Sy2s, 26.4\% for Sy1s in the
$E_{B-V}^{b}\in[0.6, 1]$ group and 44.6\% for Sy1s in the
$E_{B-V}^{b}\in[0, 0.2]$ group. These values are almost the same as
those presented in \S3.

Let's consider the fact that there are substantial Sy1s with very
low [N\,II]/H$\alpha$ first. Low [N\,II]/H$\alpha$ can be produced
with very metal poor gas (Groves et al. 2006), or presence of gas
with a density above the [N\,II] critical density with high
ionization. We argue that metal-poor gas hypothesis is unlikely.
Groves et al. showed that gas metallicity is correlated with the
mass of galaxy bulge, which is in turn correlated with central black
hole mass. However, we find that Sy1s on the left of S12 line have a
similar black hole mass distribution as the whole sample.
Furthermore, if the metal-poor gas hypothesis is right, then the
fact that this group almost does not have type 2 counterparts is
directly in contrast to the well-supported unification scheme.

The different distribution of Sy1s and Sy2s on BPT diagram does not
suggest the failure of AGN unification because reddened Sy1s show a
distribution in between. Rather it can be interpreted as partial
obscuration to the NLR in Sy2s. It is well established that the NLR
is stratified with high density and high ionization gas at close to
the continuum source whereas low density and low ionization gas is
in outer part of NLR(Nagao et al. 2003, Riffel et al. 2006). At high
densities, lines with low critical densities, such as [N\,II],
[S\,II], are suppressed by collisional de-excitation process, while
lines with high critical densities and recombination lines, are
unaffected. Considering [O\,III] is the dominant narrow line in
Sy1s, we hypothesis that even in the inner NLR, the gas density does
not exceed the critical density of [O\,III], i.e., [O\,III],
H$\alpha$ and H$\beta$ trace each other in the inner NLR. Once the
inner NLR is obscured by opaque dusty material in Sy2s, one would
expect that the H$\alpha$ emitting region to be shielded while
[N\,II] emitting region being affected to a less degree or not at
all, thus the [N\,II]/H$\alpha$ ratio will be shifted to a larger
value while [O\,III]/H$\beta$ remains the same. The uncorrected
[O\,III] luminosity of reddened Sy1s are underluminous in comparison
with unreddened one is consistent with the partial obscuration
interpretation \footnote{If the obscuring material is opaque and
covers only the inner part of NLR, narrow line flux will be smaller.
Meanwhile, due to its small size, the BLR is entirely covered by
dust and the broad lines can be efficiently corrected from Balmer
decrement.}.

The obscuring material can be the extended part of the torus or a
dusty lane in the host galaxy. The inner edge of dusty torus is
known to be order of parsecs from the central engine (Rhee et al.
2006), but the extent and height of the torus are not well
constrained. Schmitt et al. (2003) showed that all Sy1s have a
resolved bright [O\,III] emission knot at the nucleus on the scales
typically tens parsecs, and half [O\,III] emission radius is
typically tens parsecs. Thus, combined with their result, our
findings here suggest that, at least in a large fraction ($\gtrsim
30$ per cent) of AGNs if not all, the obscuring material (likely
being the torus) must extend to several tens of parsecs.

\acknowledgements  We thank Guinevere Kauffmann and Xue-Guang Zhang,
as well as the anonymous referee for critical comments. We also
thank Ting Xiao and Shaohua Zhang for useful discussion. This work
is supported by Chinese NSF grants NSF-10533050 and NSF-10573015,
the Knowledge Innovation Program (Grant No. KJCX2-YW-T05). Funding
for the Sloan Digital Sky Survey (SDSS) has been provided by the
Alfred P. Sloan Foundation, the Participating Institutions, the
National Aeronautics and Space Administration, the National Science
Foundation, the U.S. Department of Energy, the Japanese
Monbukagakusho, and the Max Planck Society. The SDSS is managed by
the Astrophysical Research Consortium (ARC) for the Participating
Institutions. The SDSS Web site is http://www.sdss.org/.

\clearpage %
\begin{figure*}
\begin{center}
\label{fig-1}
\includegraphics[width=16cm]{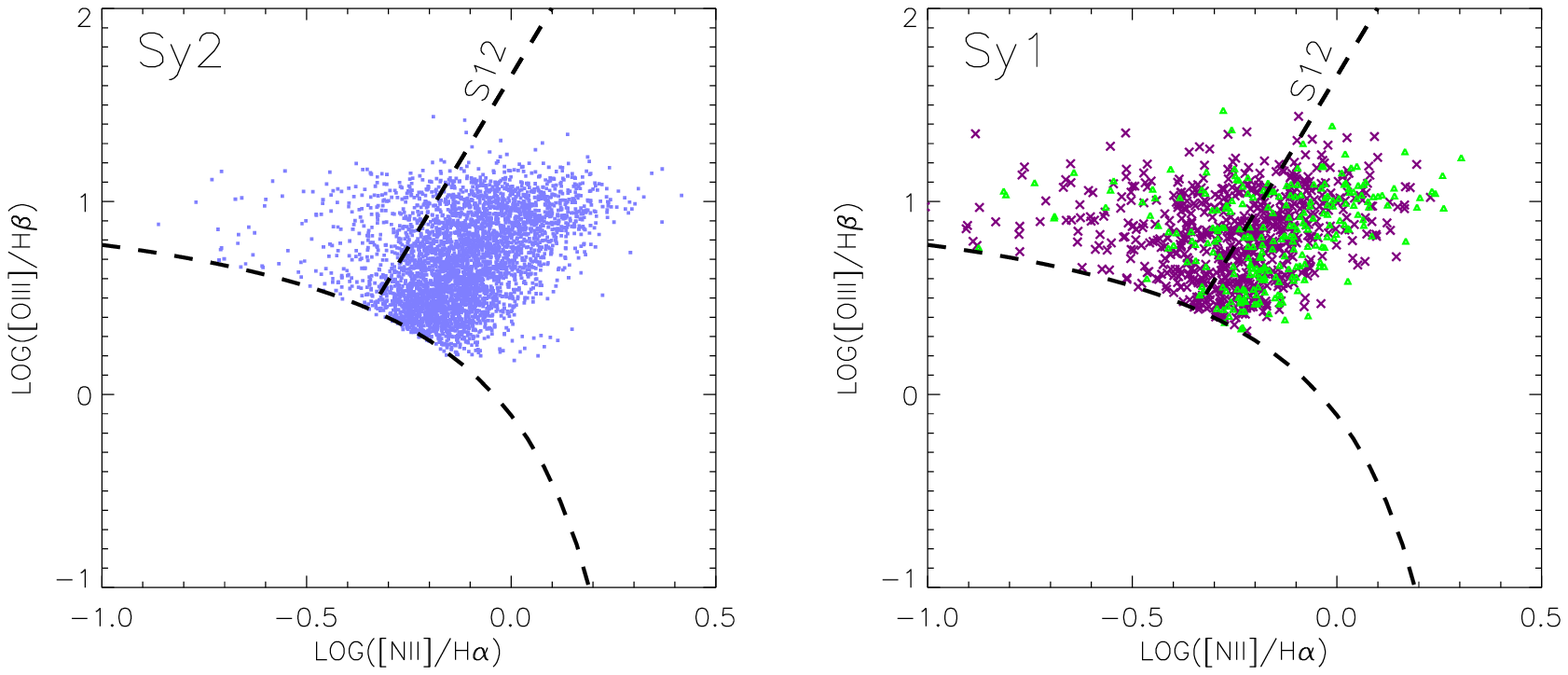}
\caption{BPT diagram for Seyfert 1 (right panel) and Seyfert 2 (left
panel) galaxies. The lower curve is the empirical line separating
AGN from star-forming galaxies (Kewley et al 2006). The straight
line is the S12 line described in the text. Most Seyfert 2 galaxies
locate on the right side of the line.  In the right panel, purple
crosses represent objects with $E_{B-V}^{b}<0.2$ and green triangles
those with $E_{B-V}^{b}\in [0.6,1]$ while the intermediate
$E_{B-V}^{b}$ group are not plotted for clarity.}

\end{center}
\end{figure*}

\begin{figure*}
\begin{center}
\label{fig-2}
\includegraphics[width=10cm]{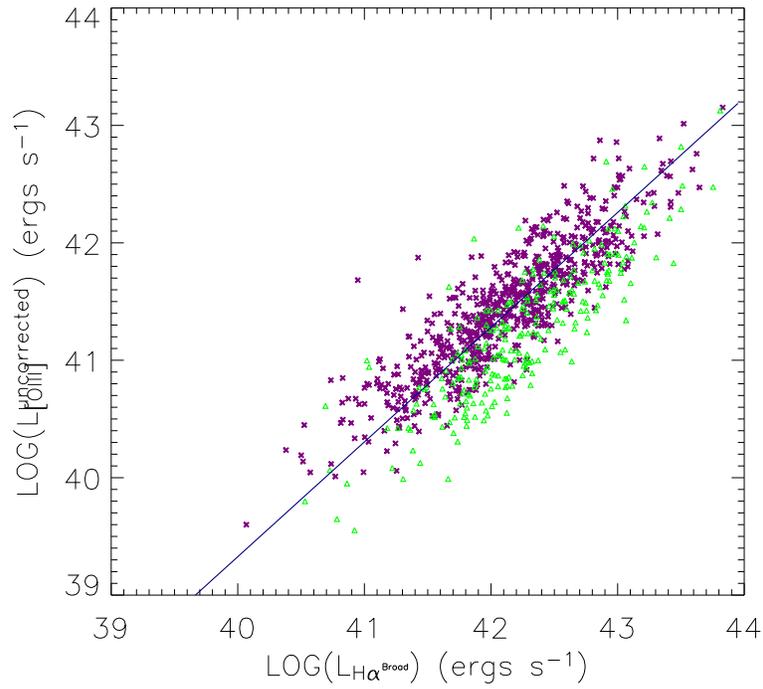}
\caption{ The uncorrected luminosity of [O\,III] versus extinction
corrected luminosity of broad H$\,\alpha$ for Seyfert 1 galaxies.
The purple crosses are AGNs with $E_{B-V}^{b}<0.2$ and green
triangles with $E_{B-V}^{b}\in [0.6,1]$. The blue line shows the
best linear fit to the whole sample. The intermediate $E_{B-V}^{b}$
group is not plotted for clarity. }
\end{center}
\end{figure*}


\end{document}